\documentclass[10pt]{article}

\usepackage{amsmath,amsfonts,amssymb,graphicx,epsfig,psfrag}
\newcommand {\I}      {\'{\i}}
\newcommand {\s}      {\sigma}

\newcommand {\Kp}     {{\mathbb K}}
\newcommand {\lp}     {{\rm 1\hskip-0.09cm l}}
\newcommand {\lpe}    {{\rm 1\hskip-0.07cm l}}
\newcommand {\sump}   {\sideset{}{^{\prime\prime}}\sum}

\newcommand {\bra}   {\langle}
\newcommand {\ket}   {\rangle_{\hskip -0.05cm\rm }}
\newcommand {\ketg}  {\rangle_{\hskip -0.05cm\rm g}}
\newcommand           {\mbe}{\left.\begin{array}}
\newcommand           {\mee}{\end{array}\right.}
\begin{document}

\title{ High Temperature Expansion for a Chain Model}
\author{Onofre Rojas$^1$, S. M. de Souza$^2$ and M. T. Thomaz$^1$\vspace{0.25cm}\\\small\it $^1$Instituto de F\I sica, Universidade Federal Fluminense
\\\small\it Av. Gal. Milton Tavares de Souza s/n.$\!\!^\circ$, CEP: 24210-340, Niter\'oi-RJ, Brazil
\vspace{0.25cm}
\\\small\it $^2$Departamento de Ci\^encias Exatas, Universidade Federal de Lavras\\\small\it Caixa Postal, 37 CEP: 37200-000, Lavras-MG,  Brazil}
\maketitle
\begin{abstract}
We consider an arbitrary translationally invariant chain model with  nearest neighbors interaction and satisfying periodic boundary condition. The approach developed here allows a thermodynamic description of the chain model directly in terms of grand potential per site. This thermodynamic function is derived from an auxiliary function constructed only from open connected sub-chains. In order to exemplify its application and how this approach works we consider the Heisenberg XXZ model. We obtain the coefficients of the high temperature expansion of the free energy per site of the model up to third order.
\end{abstract}
\vskip .5cm
\baselineskip=14pt
\section{Introduction}

Chain models, such as the spin$\frac{1}{2}$ Heisenberg models XYZ and XXZ\cite{gaudin,luther} and their extensions\cite{kirillov,alcaraz}, the one-dimensional Hubbard model\cite{lieb} etc..., have attracted intense interest due to their property of integrability.  We also have other models like the one-dimensional {\it d-p} model\cite{sanoo_pr} and its generalizations\cite{sanoo_jpsj,arracheaag}, employed to simulate a {\it Cu-O} linear chain with strong Coulomb repulsion.  These models are also relevant because they may explain high temperature superconductivity.

The solution of the XXZ model\cite{yangyang} and the one-dimensional Hubbard model\cite{lieb} at $T=0$ are examples of the success of the Bethe ansatz.  In 1972 Takahashi and Suzuki\cite{takahashi_sk} obtained the exact thermodynamics of the XXZ model. More recently the same results were re-obtained by Kuniba {\it et al.}\cite{kuniba} using the quantum transfer matrix approach. The exact thermodynamics of the one-dimensional Hubbard model was also derived by this same method\cite{bariev,ramos,juttner}. In all those models the thermodynamic functions are solutions of integral coupled equations, and it is possible to obtain their behavior in the very high temperature limit. However, it is not an easy task to derive a systematic temperature expansion for the thermodynamic functions in such temperature range.

The high temperature limit of any model can be obtained by a standard expansion in terms of the inverse of temperature $\beta$\cite{stanly}, $\beta=\frac{1}{kT}$, where $k$ is the Boltzmann constant and $T$ is the absolute temperature.  Sometimes, perturbation theory can be applied to obtain the coefficients of the $\beta$-expansion method; this approach is also called the high temperature expansion and it has been applied to the {\it t-J} model\cite{singh} and to the Hubbard model\cite{expm1d,henderson, bartkowiak}. In all standard high temperature expansions we always need to calculate the weight of each graph in the grand canonical partition function as well as to determine all graphs that contribute at a given $\beta$-order term.

Recently, we applied the basic properties of the Grassmann generators to calculate analytically the coefficients of the $\beta$-expansion of the grand canonical partition function ${\cal Z}(\beta,\mu)$ of self-interacting fermionic fields in any dimension\cite{irazietjmp}. We applied it to the one-dimensional Hubbard model\cite{irazietphya,iraziet_sub}. We have two drawbacks in our method: a large number of graphs must be calculated, as well as to count the number of times each graph contributes to each $\beta$-term in the expansion of ${\cal Z}(\beta,\mu)$.

In this paper we restrict ourselves to the study of translationally  invariant chain models with first neighbors interactions and subject to periodic boundary condition. Instead of calculating the grand canonical partition function, we derive a $\beta$-expansion for the grand potential per site, that is an intensive quantity. The coefficient of order $\beta^ n$ is exactly obtained for arbitrary values of $n$. The proposed approach is not restricted to self-interacting fermionic models.

In section 2, we expand the grand canonical partition function as a Taylor series around $\beta=0$. In fact, the idea behind  this method was previously developed, up to order $\beta^5$, in reference \cite{irazietjmp} for fermions only. Here we generalize the expansion for fermionic and bosonic fields to arbitrary orders of $\beta^n$. In performing this expansion we  write ${\cal Z}(\beta,\mu)$ as a sum of  traces whose coefficients are known.  We obtain the expansion of the grand potential per site, and verify that it is an intensive quantity. In this calculation we do not use the traditional method of quantum transfer matrix.  As an illustrative example  of the method developed here, in section 3 we apply it to spin-$\frac{1}{2}$ Heisenberg XXZ model. We calculate the $\beta$-expansion   of the Helmholtz free energy of the model up to order $\beta^3$. Taking $\Delta=h=0$ we recover the $\beta$-expansion of this thermodynamic function of the free fermion model\cite{klumper93} whereas when $t=0$ we recover the limiting Ising model\cite{L_baxter}. In section 4 we draw our main conclusions. In Appendix A we explain our graphic representation of the normalized traces; in Appendix B we extend our results for $\bra\Kp^n\ket$ with arbitrary $n$; in Appendix C we write the grand potential per site in terms of an auxiliary function $\varphi(\lambda)$ that is constructed only from open connected sub-chains; and finally in Appendix D we write the function ${\rm K}_{1,m}^{(n)}$ as sums of normalized traces for $n$ up to $4$.

\section{Grand Canonical Partition Function for a Chain Model}
Let us consider a one-dimensional regular lattice (a periodic chain) with $N$ sites, so that the Hilbert space of the chain model is simply ${\mathcal H}^{(N)}=\overset{N}\otimes{\mathcal H}$, ${\mathcal H}$ being the irreducible representation at one site, including all its degrees to freedom. The dimension of this Hilbert space is ${\rm dim}{\mathcal H}^{(N)}={\rm tr}_N(\lp)$. The notation ${\rm tr}_N$ means the trace over all $N$ sites and their internal degrees of freedom, such e.g. spin. 

The grand canonical partition function of a quantum system in the chain with $N$ sites is given by
\begin{equation}
{\mathcal Z}_N(\beta,\mu)={\rm tr}_N({\rm e}^{-\beta\Kp}),
\end{equation}
where  $\Kp={\mathbb H}-\mu{\mathbb N}$, with  $\mu$ being the chemical potential and ${\mathbb N}$ being an operator that acts on ${\mathcal H}^{(N)}$ and that commutes with the Hamiltonian of the system. The expansion of ${\cal Z}_N(\beta,\mu)$ around $\beta=0$ is
\begin{equation}\label{tyl K}
{\cal Z}_N(\beta,\mu)={\rm tr}_N(\lp)+\sum_{n=1}^{\infty}(-\beta)^{n}\frac{{\rm tr}_N(\Kp^n)}{n!}.
\end{equation}

Let  ${\bf A}$ be any operator that acts on ${\mathcal H}^{(M)}$ where $M\leqslant N$. We define $\bra {\bf A}\ket\equiv\frac{{\rm tr}_M({\bf A})}{{\rm tr}_M(\lp)}$, for any dimension of ${\mathcal H}^{(M)}$.  From now on, we call $\bra {\bf A}\ket$ the {\it normalized trace} of operator ${\bf A}$. The dimension of the subspace ${\mathcal H}^{(M)}$ is determined by the operator ${\bf A}$. 

Using the definition of normalized trace, eq.(\ref{tyl K}) becomes

\begin{equation}\label{tyl K1}
{\cal Z}_N(\beta,\mu)= {\rm tr}_N(\lp)\Big\{1+\sum_{n=1}^{\infty}(-\beta)^{n}\frac{\bra\Kp^n\ket}{n!}\Big\}.
\end{equation}

Along this section we consider a general Hamiltonian ${\mathbb H}$ subject to two constraints: the interaction is only between first neighbors and the Hamiltonian ${\mathbb H}$ is invariant under translation along the chain.  The most general operator $\Kp$ satisfying both conditions is
\begin{equation}\label{def K}
{\mathbb K}= \sum_{i=1}^{N}\widetilde{\bf K}_{i,i+1},
\end{equation}
where $\widetilde{\bf K}_{i,i+1}\in {\mathcal H}^{(N)}$.  Each operator $\widetilde{\bf K}_{i,i+1}$ is defined as
\begin{align}
\widetilde{\bf K}_{1,2}&= {\bf K}_{1,2}\otimes{\bf 1}_3\otimes\dots\otimes{\bf 1}_N\nonumber\\ \widetilde{\bf K}_{2,3}&= {\bf 1}_1\otimes {\bf K}_{2,3}\otimes{\bf 1}_4\otimes\dots\otimes{\bf 1}_N\nonumber\\ \vdots&\nonumber\\ \widetilde{\bf K}_{i,i+1}&= {\bf 1}_1\otimes\dots\otimes{\bf 1}_{i-1}\otimes {\bf K}_{i,i+1}\otimes{\bf 1}_{i+2}\otimes\dots\otimes{\bf 1}_{N}
\end{align}
and ${\bf K}_{i,i+1}\in {\mathcal H}_i\otimes{\mathcal H}_{i+1}$. We use the notation ${\bf 1}_i\in {\mathcal H}_i$ for the identity matrix on the irreducible sub-space of the $i-{\rm th}$ particle.

Our aim is to get the coefficients $\bra{\mathbb K}^n\ket$, on the r.h.s. of the eq.(\ref{tyl K1}), in terms of the normalized traces of products of operators ${\bf K}_{i,i+1}$. Let us calculate explicitly the first three coefficients ($n=1,\;n=2$, and $n=3$) of the expansion (\ref{tyl K1}). They will help us to construct the coefficient $\bra\Kp^n\ket$ for arbitrary $n$. We begin with $n=1$.

 Since the operator ${\bf K}_{i,i+1}$ acts only on the sites $i$ and $i+1$, we have 
\begin{align}\label{vmK12} 
{\rm tr}_N(\widetilde{\bf K}_{i,i+1}) ={\rm tr}_2({\bf K}_{i,i+1})\big({\rm tr}_1({\bf 1})\big)^{N-2}.
\end{align}
We stress out that the traces of the r.h.s. of eq.(\ref{vmK12}) are calculated on the subspace ${\mathcal H}^{(1)}$ and  ${\mathcal H}^{(2)}$ whereas the trace on the l.h.s. is calculated on the complete space ${\mathcal H}^{(N)}$. Dividing both sides of eq.(\ref{vmK12}) by ${\rm tr}_N(\lpe)$, 
\begin{equation}\label{nvmK12} 
\frac{{\rm tr}_N(\widetilde{\bf K}_{i,i+1})}{{\rm tr}_N(\lpe)} =\frac{{\rm tr}_2({\bf K}_{i,i+1})}{{\rm tr}_2({\bf 1}\otimes{\bf 1})}=\bra {\bf K}_{i,i+1}\ket.
\end{equation}

Due to the property of the Hamiltonian being translationally invariant we have: $\bra\widetilde{\bf K}_{i,i+1}\ket=\bra {\bf K}_{i,i+1}\ket=\bra {\bf K}_{1,2}\ket$. 
Taking into account the periodic boundary condition (${\bf K}_{N,N+1}={\bf K}_{N,1}$), we have that $\bra{\mathbb K}\ket$ is equal to
\begin{equation}\label{K1}
\bra{\mathbb K}\ket =\sum_{i=1}^N\bra{\bf K}_{i,i+1}\ket =N\bra{\bf K}_{1,2}\ket.
\end{equation}
 For $n=2$ on the r.h.s. of expansion (\ref{tyl K1}) we have to calculate the normalized traces of $\bra{\mathbb K}^2\ket$. From the definition of operator ${\mathbb K}$ (see eq.(\ref{def K})), we have
\begin{equation}\label{sum K2}
\bra{\mathbb K}^2\ket =\sum_{i,j=1}^N\bra{\bf K}_{i,i+1}{\bf K}_{j,j+1}\ket.
\end{equation}
 In the normalized trace $\bra {\bf K}_{i,i+1}{\bf K}_{j,j+1}\ket$ we have three different cases ( in Appendix A we explain our graphic representation):
\begin{subequations}
\begin{itemize}
\item[{\it i})] $i=j$:
\begin{eqnarray}\label{K112}
\bra{\bf K}_{i,i+1}^2\ket=\bra{\bf K}_{1,2}^2\ket
\quad\longrightarrow\quad
\begin{picture}(100,10)
\thicklines
\put(40,2){\line(1,0){30}}
\put(40,12){\line(1,0){30}}
\multiput(40,12)(30,0){2}{\circle{1}}
\multiput(40,12)(30,0){2}{\circle{2}}
\multiput(0,2)(4,0){30}{\line(1,0){1}}
\multiput(40,2)(0,2){5}{\line(0,1){1}}
\multiput(70,2)(0,2){5}{\line(0,1){1}}
\multiput(10,2)(30,0){4}{\circle{1}}
\multiput(10,2)(30,0){4}{\circle{2}}
\put(0,-8){\footnotesize{$i-1$}}
\put(38,-8){\footnotesize{$i$}}
\put(60,-8){\footnotesize{$i+1$}}
\put(94,-8){\footnotesize{$i+2$}}
\end{picture}
\end{eqnarray}
This equality comes from the invariance under translation and the periodic boundary condition satisfied by the model. In the double sum (\ref{sum K2}) we have $N$ terms of this type.
\item[{\it ii})] $i=j\pm 1$:
\begin{align}\label{j+1}
\bullet&&  i=j+1:&&
\bra{\bf K}_{i,i+1}{\bf K}_{i+1,i+2}\ket&&
\longrightarrow&&
\begin{picture}(100,10)
\thicklines
\put(10,2){\line(1,0){30}}
\put(40,12){\line(1,0){30}}
\multiput(40,12)(30,0){2}{\circle{1}}
\multiput(40,12)(30,0){2}{\circle{2}}
\multiput(0,2)(4,0){30}{\line(1,0){1}}
\multiput(40,2)(0,2){5}{\line(0,1){1}}
\multiput(70,2)(0,2){5}{\line(0,1){1}}
\multiput(10,2)(30,0){4}{\circle{1}}
\multiput(10,2)(30,0){4}{\circle{2}}
\put(10,-8){\footnotesize{$i$}}
\put(30,-8){\footnotesize{$i+1$}}
\put(62,-8){\footnotesize{$i+2$}}
\put(92,-8){\footnotesize{$i+3$}}
\end{picture}&&
\\
\label{j-1}
\bullet&& i=j-1:&&
\bra{\bf K}_{i+1,i+2}{\bf K}_{i,i+1}\ket&&
\longrightarrow&&
\begin{picture}(100,30)
\thicklines
\put(40,2){\line(1,0){30}}
\put(10,12){\line(1,0){30}}
\multiput(10,12)(30,0){2}{\circle{1}}
\multiput(10,12)(30,0){2}{\circle{2}}
\multiput(0,2)(4,0){30}{\line(1,0){1}}
\multiput(10,2)(0,2){5}{\line(0,1){1}}
\multiput(40,2)(0,2){5}{\line(0,1){1}}
\multiput(10,2)(30,0){4}{\circle{1}}
\multiput(10,2)(30,0){4}{\circle{2}}
\put(10,-8){\footnotesize{$i$}}
\put(30,-8){\footnotesize{$i+1$}}
\put(62,-8){\footnotesize{$i+2$}}
\put(92,-8){\footnotesize{$i+3$}}
\end{picture}&&
\end{align}

In the double sum on the r.h.s. of eq.(\ref{sum K2}) we have $N$ terms of the type of one of these two previous configurations. Using the cyclic property of traces and  translation invariance, we get that the normalized traces (\ref{j+1}) and (\ref{j-1}) are equal. Therefore, we can write: $\bra{\bf K}_{i,i+1}{\bf K}_{i-1,i}\ket=\bra{\bf K}_{i,i+1}{\bf K}_{i+1,i+2}\ket$.

\item [{\it iii})] $i\ne j$ and $i\ne j\pm 1$:
\begin{equation}\label{k12 k34}
\bra{\bf K}_{i,i+1}{\bf K}_{j,j+1}\ket
\qquad\longrightarrow\qquad
\begin{picture}(100,20)
\thicklines
\put(10,2){\line(1,0){30}}
\put(70,12){\line(1,0){30}}
\multiput(70,12)(30,0){2}{\circle{1}}
\multiput(70,12)(30,0){2}{\circle{2}}
\multiput(0,2)(4,0){30}{\line(1,0){1}}
\multiput(70,2)(0,2){5}{\line(0,1){1}}
\multiput(100,2)(0,2){5}{\line(0,1){1}}
\multiput(10,2)(30,0){4}{\circle{1}}
\multiput(10,2)(30,0){4}{\circle{2}}
\put(8,-8){\footnotesize{$i$}}
\put(30,-8){\footnotesize{$i+1$}}
\put(68,-8){\footnotesize{$j$}}
\put(95,-8){\footnotesize{$j+1$}}
\end{picture}
\end{equation}
The normalized trace (\ref{k12 k34}) can be written as
$\bra{\bf K}_{i,i+1}\ket\bra{\bf K}_{j,j+1}\ket=\bra{\bf K}_{1,2}\ket^2$. The number of terms in the double sum in eq.(\ref{sum K2}) that satisfies configuration (\ref{k12 k34}) is equal to $N(N-3)$.
\end{itemize}
\end{subequations}
The  terms that contribute to $\bra{\mathbb K}^2\ket$, with respective of number of configurations (weight) in the sum on the r.h.s. of eq.(\ref{sum K2}) are
\begin{equation}\label{K2}
\bra\Kp^2\ket = N\bra{\bf K}_{1,2}^2\ket+2N\bra{\bf K}_{1,2}{\bf K}_{2,3}\ket+N(N-3)\bra{\bf K}_{1,2}{\bf K}_{3,4}\ket.
\end{equation}
The last term on the r.h.s of (\ref{K2}) can be written as product of normalized traces with $n=1$.

 The term $\bra{\mathbb K}^3\ket$ have a richer structure of traces than the two previous ones. Its explicit study will help us to generalize the results for arbitrary $n$. The expansion of normalized traces in $\bra{\mathbb K}^3\ket$ is
\begin{equation}\label{sum_K3}
\bra{\mathbb K}^3\ket =\sum_{i,j,k=1}^N\bra{\bf K}_{i,i+1}{\bf K}_{j,j+1}{\bf K}_{k,k+1}\ket.
\end{equation}

Similarly to $\bra{\mathbb K}\ket$ and $\bra{\mathbb K}^2\ket$, we determine the types of configurations we have in the sum on the r.h.s of eq.(\ref{sum_K3}) and their respectively weight. For $\bra{\mathbb K}^3\ket$ we have
\begin{align}\label{K3}
\bra\Kp^3\ket=& N\bra{\bf K}_{1,2}^3\ket+3N\Big\{\frac{\bra{\bf K}_{1,2}^2{\bf K}_{2,3}\ket}{2!}+\frac{\bra{\bf K}_{1,2}{\bf K}_{2,3}^2\ket}{2!}\Big\}+3N \big\{\bra{\bf K}_{1,2}{\bf K}_{2,3}{\bf K}_{3,4}\ket+\bra{\bf K}_{1,2}{\bf K}_{3,4}{\bf K}_{2,3}\ket\big\}+\nonumber\\& +3N(N-3)\Big\{\frac{\bra{\bf K}_{1,2}^2{\bf K}_{3,4}\ket}{2!}+\frac{\bra{\bf K}_{1,2}{\bf K}_{3,4}^2\ket}{2!}\Big\}+3N(N-4)\big\{\bra{\bf K}_{1,2}{\bf K}_{2,3}{\bf K}_{4,5}\ket+\nonumber\\& +\bra{\bf K}_{1,2}{\bf K}_{3,4}{\bf K}_{4,5}\ket\big\}+N(N-4)(N-5)\bra{\bf K}_{1,2}{\bf K}_{3,4}{\bf K}_{5,6}\ket.
\end{align}
From the expressions (\ref{K1}), (\ref{K2}) and (\ref{K3}) we have that each type of normalized trace (in the expression of $\bra{\mathbb K}^3\ket$ we collect them inside of braces) corresponds to all normalized traces that we can write for that set of operators. For example, the term: $3\bra{\bf K}_{1,2}{\bf K}_{2,3}{\bf K}_{3,4}\ket+3\bra{\bf K}_{1,2}{\bf K}_{3,4}{\bf K}_{2,3}\ket$ includes all normalized traces that we can write for the set of operators $\{{\bf K}_{1,2}\;,{\bf K}_{2,3}\;,{\bf K}_{3,4}\}$. The factors $3$ comes from the fact that some of them are equals by the cyclic property of the trace.

To make our notation simpler and generalize our results for arbitrary $n$, we define the {\it g-trace}:
\begin{align}\label{def Kg}
\left\bra{\bf K}_{{ i_1,i_1+1}}^{n_1}{\bf K}_{i_2,i_2+1}^{n_2}\dots{\bf K}_{i_m,i_m+1}^{n_m}\right\ketg\equiv\frac{n_1!\;n_2!\;\dots\; n_m!}{n!}\sum_{\mathcal P}\bra {\mathcal P}({\bf K}_{i_1,i_1+1}^{n_1}\,,{\bf K}_{i_2,i_2+1}^{n_2}\,,\dots,{\bf K}_{i_m,i_m+1}^{n_m})\ket,
\end{align}
where $\sum_{i=1}^{m}n_i=n$ with $n_i\ne 0$ and the indices $i_k$, $k=1..m$ are distinct among themselves.
 By definition, $\bra {\mathcal P}({\bf K}_{i_1,i_1+1}\,,{\bf K}_{i_2,i_2+1}\,,\dots,{\bf K}_{i_m,i_m+1})\ket$ represents all distinct permutations that we can write the $n$ operators $\{{\bf K}_{i_1,i_1+1}\,,{\bf K}_{i_2,i_2+1}\,,\dots,{\bf K}_{i_m,i_m+1}\}$. In the particular case when the operators $\{{\bf K}_{i_1,i_1+1}\,,{\bf K}_{i_2,i_2+1}\,,\dots\break \dots,{\bf K}_{i_m,i_m+1}\}$ commute among themselves, $\bra{\bf K}_{{ i_1,i_1+1}}^{n_1}{\bf K}_{i_2,i_2+1}^{n_2}\dots{\bf K}_{i_m,i_m+1}^{n_m}\ketg$ reduces to the normalized trace $\bra{\bf K}_{{ i_1,i_1+1}}^{n_1}{\bf K}_{i_2,i_2+1}^{n_2}\dots{\bf K}_{i_m,i_m+1}^{n_m}\ket$.

The normalized traces $\bra{\mathbb K}\ket$, $\bra{\mathbb K}^2\ket$ and  $\bra{\mathbb K}^3\ket$ written in terms of the g-traces become
\begin{subequations}\label{k1k2k3}
\begin{align}
\bra\Kp\ket=&N\bra{\bf K}_{1,2}\ketg,\\
\nonumber\\
\frac{\bra\Kp^2\ket}{2!}= &N\frac{\bra{\bf K}_{1,2}^2\ketg}{2!}+N\bra{\bf K}_{1,2}{\bf K}_{2,3}\ketg+\frac{N(N-3)}{2!}\bra{\bf K}_{1,2}{\bf K}_{3,4}\ketg,
\end{align}
and
\begin{align}
\frac{\bra\Kp^3\ket}{3!}= &N\frac{\bra{\bf K}_{1,2}^3\ketg}{3!}+N\Big\{\frac{\bra{\bf K}_{1,2}^2{\bf K}_{2,3}\ketg}{2!}+\frac{\bra{\bf K}_{1,2}{\bf K}_{2,3}^2\ketg}{2!}\Big\}+N \bra{\bf K}_{1,2}{\bf K}_{2,3}{\bf K}_{3,4}\ketg+\nonumber\\& +\frac{N(N-3)}{2!}\Big\{\frac{\bra{\bf K}_{1,2}^2{\bf K}_{3,4}\ketg}{2!}+\frac{\bra{\bf K}_{1,2}{\bf K}_{3,4}^2\ketg}{2!}\Big\}+\frac{N(N-4)}{2!}\big\{\bra{\bf K}_{1,2}{\bf K}_{2,3}{\bf K}_{4,5}\ketg+\nonumber\\& +\bra{\bf K}_{1,2}{\bf K}_{3,4}{\bf K}_{4,5}\ketg\big\}+\frac{N(N-4)(N-5)}{3!}\bra{\bf K}_{1,2}{\bf K}_{3,4}{\bf K}_{5,6}\ketg.
\end{align}
\end{subequations}

Some of the g-traces at order $n$  can be written as product of the normalized traces of lower order. The latter  are already calculated, which means that we have a smaller number of new normalized traces to be calculated at order $n$. To distinguish the normalized traces that appear for the first time at order $n$ from the others we use the notation ${\rm K}_{r,m}^{(n)}$, where $n$ is the power  $(\bra{\mathbb K}^n\ket)$ of the term in expansion (\ref{tyl K1}) to which the g-traces contribute, $r$ is the number of products of the traces that the original trace can be broken and $m$ is the number of distinct ${\bf K}_{i,i+1}$ operators that appear in the g-trace. The terms  ${\rm K}_{r,m}^{(n)}$ with $r=1$ are the g-traces that appear in the expansion for the first time at order $\beta^n$.

We begin by defining ${\rm K}_{1,m}^{(n)}$ as
\begin{subequations}\label{k1mn}
\begin{align}
{\rm K}_{1,1}^{(n)}&=\frac{\bra{\bf K}_{1,2}^n\ketg}{n!},\label{K11n}\\
{\rm K}_{1,2}^{(n)}&={\sump_{\{n_i\}}^{n}}\frac{\bra{\bf K}_{1,2}^{n_1}{\bf K}_{2,3}^{n_2}\ketg}{n_1!\;n_2!\;\;},\\
&\vdots\nonumber\\
{\rm K}_{1,m}^{(n)}&={\sump_{\{n_i\}}^{n}}\frac{\bra{\bf K}_{1,2}^{n_1}{\bf K}_{2,3}^{n_2}\dots{\bf K}_{m,m+1}^{n_m}\ketg}{n_1!\;n_2!\;\dots\; n_m!\;\;\;}.\label{K1mn}
\end{align}
\end{subequations}
 From now on we use the notation {\scriptsize$\underset{\{n_i\}}{\overset{n}{\sump}}$} to mean the restriction: $\underset{i=1}{\overset{n}{\sum}}n_i=n$ and $n_i\ne 0$ for $i=1,2,..,m$. The index $m$ satisfies the condition: $1\leqslant m \leqslant{\rm min}(n,N)$.  In eq.(\ref{K1mn}) we are assuming $m\leqslant n$. However, if $m>n$ we define ${\rm K}_{1,m}^{(n)}=0$.

 Since in the definition of the g-traces are included all possible permutation of the $n$ operators, we have
\begin{align}
{\rm K}_{1,m}^{(n)}={\sump_{\{n_i\}}^{n}}\big\bra\prod_{i=1}^{m}\frac{{\bf K}_{i,i+1}^{n_i}}{n_i!}\big\ketg.
\end{align}
For $r=2$, the natural definition is 
\begin{subequations}
\begin{align}
{\rm K}_{2,2}^{(n)}&={\sump_{\{n_i\}}^{n}}\frac{\bra{\bf K}_{1,2}^{n_1}{\bf K}_{3,4}^{n_2}\ketg}{n_1!\;n_2!\;\;}={\sump_{\{n_i\}}^{n}}\frac{\bra{\bf K}_{1,2}^{n_1}\ketg}{n_1!}\frac{\bra{\bf K}_{3,4}^{n_2}\ketg}{n_2!\;}={\sump_{\{n_i\}}^{n}}{\rm K}_{1,1}^{(n_1)}{\rm K}_{1,1}^{(n_2)}\;,\label{K22n}\\ {\rm K}_{2,3}^{(n)}&={\sump_{\{n_i\}}^{n}}\left(\frac{\bra{\bf K}_{1,2}^{n_1}{\bf K}_{2,3}^{n_2}{\bf K}_{4,5}^{n_3}\ketg}{n_1!\;n_2!\;n_3!\;\;}+\frac{\bra{\bf K}_{1,2}^{n_1}{\bf K}_{3,4}^{n_2}{\bf K}_{4,5}^{n_3}\ketg}{n_1!\;n_2!\;n_3!\;\;}\right)={\sump_{\{n_i\}}^{n}}\left({\rm K}_{1,1}^{(n_1)}{\rm K}_{1,2}^{(n_2)}+{\rm K}_{1,2}^{(n_1)}{\rm K}_{1,1}^{(n_2)}\right),\label{K23n}
\end{align}
\end{subequations}
where we took into account the invariance of the model under translations and the fact that
\begin{align}
[{\bf K}_{i,i+1},{\bf K}_{j,j+1}]=\begin{cases}{\bf 0}& \text{when $i\ne j\pm 1$},\\ \ne{\bf 0} & \text{when $i=j\pm 1$}.
\end{cases}
\end{align}
To obtain the second equality on the r.h.s. of eq.(\ref{K23n}) we  rearrange the indices in the double sum.
We see that these g-traces can always be writing as product of g-traces of lower order which give us
\begin{equation}\label{K2mn}
{\rm K}_{2,m}^{(n)}={\sump_{\{n_i\}}^{n}}\sum_{l=1}^{m-1}{\rm K}_{1,l}^{(n_1)}{\rm K}_{1,m-l}^{(n_2)}={\sump_{\{n_i\}}^{n}}{\sump_{\{m_j\}}^{m}}{\rm K}_{1,m_1}^{(n_1)}{\rm K}_{1,m_2}^{(n_2)}\;.
\end{equation}
The indices $n_i$ satisfy the condition $n_1+n_2=n$ while $m_1+m_2=m$ and $2\leqslant m \leqslant{\rm min}(n,N)$.  We are assuming $m\leqslant n$, otherwise ${\rm K}_{2,m}^{(n)}=0$. For ${\rm K}_{3,3}^{(n)}$ and ${\rm K}_{3,4}^{(n)}$ we have
\begin{subequations}
\begin{align}
{\rm K}_{3,3}^{(n)}&={\sump_{\{n_i\}}^{n}}{\rm K}_{1,1}^{(n_1)}{\rm K}_{1,1}^{(n_2)}{\rm K}_{1,1}^{(n_3)}\;,\\ {\rm K}_{3,4}^{(n)}&={\sump_{\{n_i\}}^{n}}\left({\rm K}_{1,1}^{(n_1)}{\rm K}_{1,1}^{(n_2)}{\rm K}_{1,2}^{(n_3)}+{\rm K}_{1,1}^{(n_1)}{\rm K}_{1,2}^{(n_2)}{\rm K}_{1,1}^{(n_3)}+{\rm K}_{1,2}^{(n_1)}{\rm K}_{1,1}^{(n_2)}{\rm K}_{1,1}^{(n_3)} \right).
\end{align}
\end{subequations}
It follows for arbitrary $m$, where $3\leqslant m\leqslant{\rm min}(n,N)$,
\begin{align}
{\rm K}_{3,m}^{(n)}&= {\sump_{\{n_i\}}^{n}}{\sump_{\{m_i\}}^{m}}{\rm K}_{1,m_1}^{(n_1)}{\rm K}_{1,m_2}^{(n_2)}{\rm K}_{1,m_3}^{(n_3)}\;,\label{K3mn}
\end{align}
and ${\rm K}_{3,m}^{(n)}=0$ for $m>n$.
  
We generalize our definition of ${\rm K}_{r,m}^{(n)}$ 
\begin{equation}\label{Krmn}
{\rm K}_{r,m}^{(n)}\equiv {\sump_{\{n_i\}}^{n}} {\sump_{\{m_i\}}^{m}}\prod_{j=1}^{r}{\rm K}_{1,m_j}^{(n_j)}\;,
\end{equation}
where $\{n_i\}\equiv\{n_1,n_2,\dots,n_r\}$ and $\{m_i\}\equiv\{m_1,m_2,\dots,m_r\}$. 
Eq.(\ref{Krmn}) is valid for $r>1$ and it makes possible to write  ${\rm K}_{r,m}^{(n)}$ as products of ${\rm K}_{1,m}^{(n)}$ (see eq.(\ref{K1mn})). The indices $r$ and $m$ satisfy the conditions $1\leqslant r\leqslant{\rm min}(n,N)$ and $r\leqslant m\leqslant{\rm min}(n,N)$. We are assuming $n\geqslant r$ and $m$, otherwise ${\rm K}_{r,m}^{(n)}=0$.

In terms of ${\rm K}_{r,m}^{(n)}$, the normalized traces $\bra\Kp\ket$, $\bra\Kp^2\ket$ and  $\bra\Kp^3\ket$ are simply
\begin{subequations}\label{K 1 2 3}
\begin{align}
\bra\Kp\ket&= N{\rm K}_{1,1}^{(1)}\;,\\
\frac{\bra\Kp^2\ket}{2!}&= N{\rm K}_{1,1}^{(2)}+N{\rm K}_{1,2}^{(2)}+\frac{N(N-3)}{2!}{\rm K}_{2,2}^{(2)}\;,\\
\frac{\bra\Kp^3\ket}{3!}&= N{\rm K}_{1,1}^{(3)}+N{\rm K}_{1,2}^{(3)}+N{\rm K}_{1,3}^{(3)}+\frac{N(N-3)}{2!}{\rm K}_{2,2}^{(3)}+\frac{N(N-4)}{2!}{\rm K}_{2,3}^{(3)}+\frac{N(N-4)(N-5)}{3!}{\rm K}_{3,3}^{(3)}\;.
\end{align}
\end{subequations}

In the Appendix {\bf B} we write $\bra\Kp^4\ket$ in terms of ${\rm K}_{r,m}^{(4)}$ and the coefficients of ${\rm K}_{1,m}^{(n)}$, ${\rm K}_{2,m}^{(n)}$and  ${\rm K}_{3,m}^{(n)}$ for $n=1,2$ and $3$ are rewritten in terms of binomial coefficients. In this appendix we also extend the previous results for arbitrary $n$.

 For arbitrary $n$, eq.(\ref{TrnK Ap}) gives the expression of  $\bra\Kp^n\ket$, that is
\begin{equation}\label{TrnK}
\frac{\bra{\Kp}^n\ket}{n!}=\sum_{r=1}^{[n,N]}\sum_{m=r}^{[n,N]}\frac{N}{r}\binom{N-m-1}{r-1} {\rm K}^{(n)}_{r,m}\;.
\end{equation}
The notation $[n,N]$ means the ${\rm min}(n,N)$.
 Differently from our previous work\cite{irazietjmp,irazietphya,iraziet_sub}, in eq.(\ref{TrnK}) we already have the weight of each set of sub-chain ${\rm K}_{r,m}^{(n)}$ in $\bra{\Kp}^n\ket$. The coefficient that multiplies ${\rm K}_{r,m}^{(n)}$ in eq.(\ref{TrnK}) is independent of the particular Hamiltonian under consideration.

 For $n<N$ the trace will be calculated on a maximum subspace given by ${\mathcal H}^{(n+1)}$, whereas for $n\geqslant N$ due to the periodic boundary condition the operators act on whole Hilbert space ${\mathcal H}^{(N)}$, therefore we can say that the operators act more than one period in the periodic chain.

Rewriting the coefficient of ${\rm K}_{r,m}^{(n)}$ in eq.(\ref{TrnK}) in a more convenient way,
\begin{align}\label{eq Trkc}
\frac{\bra{\Kp}^n\ket}{n!}&=\sum_{r=1}^{[n,N]}\sum_{m=r}^{[n,N]}\sum_{k=1}^r(-1)^{r+k}\frac{k}{r}\binom{m+r-k-1}{r-k}\binom{N}{k}{\rm K}_{r,m}^{(n)}\nonumber\\&\equiv\sum_{k=1}^{[n,N]}\binom{N}{k}{\mathfrak K}_{k,n}
\end{align}
with
\begin{equation}\label{k_rn}
{\mathfrak K}_{k,n}\equiv\sum_{r=k}^{[n,N]}\sum_{m=r}^{[n,N]}(-1)^{r+k}\frac{k}{r}\binom{m+r-k-1}{r-k}{\rm K}_{r,m}^{(n)}.
\end{equation}
From eq. (\ref{k_rn}) we have that the biggest possible value of $k$ is ${\rm min}(n,N)$ while there is no restriction on $n$.  
The function ${\mathfrak K}_{k,n}$ has the  property,
\begin{equation}\label{p_krn}
{\mathfrak K}_{k,n}=\sump_{\{n_i\}}^{n}\prod_{i=1}^k{\mathfrak K}_{1,n_i},
\end{equation}
where $\{n_i\}\equiv\{n_1,n_2,\dots,n_k\}$. Replacing eq.(\ref{eq Trkc}) in eq.(\ref{tyl K1}) we get
\begin{equation}\label{s_trk}
{\cal Z}_N(\beta,\mu)= {\rm tr}_N(\lp)\Big\{1+\sum_{k=1}^{N}\binom{N}{k}\sum_{n=k}^{\infty}(-\beta)^{n}{\mathfrak K}_{k,n}\Big\}.
\end{equation}
From property (\ref{p_krn}) and recombining the summations conveniently, eq.(\ref{s_trk}) is finally written for any value of $N$ as
\begin{equation}\label{F_trk}
{\cal Z}_N(\beta,\mu)=\Big\{ {\rm tr}_1({\bf 1})\big(1+\sum_{n=1}^{\infty}(-\beta)^{n}{\mathfrak K}_{1,n}\big)\Big\}^N.
\end{equation}
The grand potential per site ${\cal W}(\beta,\mu)$ is
\begin{equation}\label{Wbeta}
{\cal W}_N(\beta,\mu)=-\frac{1}{N\beta}\ln{\cal Z}_N(\beta,\mu)=-\frac{1}{\beta}\ln\Big({\rm tr}_1({\bf 1})\big(1+\sum_{n=1}^{\infty}(-\beta)^{n}{\mathfrak K}_{1,n}\big)\Big).
\end{equation}
This expansion can be used to obtain analytical results of chain models in the high temperature limit.

We can be misleading by the notation ${\mathfrak K}_{1,n}$ in eq.(\ref{Wbeta}) since in it are included not only connected sub-chains but also disconnected ones. For finite $N$, eq.(\ref{Wbeta}) is the simplest form of the grand potential per site since in this case $n$ can be bigger than $N$ which means to go around the chain more than once.

\subsection{The Thermodynamic Limit}

Our main interest is to get the thermodynamical behavior of physical quantities that characterize physical systems. Those are attained only in the thermodynamic limit ($N\rightarrow\infty$). We first take the thermodynamic limit and after makes the expansion (\ref{tyl K}), so that we never goes along the chain more than once when we calculate the normalized traces. In this thermodynamic limit the r.h.s. of eq.(\ref{F_trk}) is equal to the highest eigenvalue obtained from the  associated quantum transfer matrix\cite{L_baxter} to the quantum system of interest. From eq.(\ref{Wbeta}), the grand potential per site ${\cal W}(\beta,\mu)$ in this limit is
\begin{equation}\label{Wbt}
{\cal W}(\beta,\mu)=-\lim_{N\rightarrow \infty}\frac{1}{N\beta}\ln{\cal Z}_N(\beta,\mu)=-\frac{1}{\beta}\ln\Big({\rm tr}_1({\bf 1})\big(1+\sum_{n=1}^{\infty}(-\beta)^{n}{\mathfrak K}_{1,n}\big)\Big).
\end{equation}
Eqs.(\ref{Wbeta}) and (\ref{Wbt}), look alike. Their difference relies on the fact that in eq.(\ref{Wbt}) the biggest sub-chain that contribute to ${\mathfrak K}_{1,n}$ has at most the length of the chain. Eq.(\ref{Wbt}) confirms that the grand potential is an intensive quantity.

Certainly our aim is to be able to calculate the smallest possible number of open sub-chains to obtain the grand potential per site. But the definition of ${\mathfrak K}_{1,n}$ (see eq.(\ref{k_rn})) includes connected as well disconnected sub-chains.
 In eq.(\ref{Wbt}) we have the following sum to calculate
\begin{equation}
\xi\equiv\sum_{n=1}^{\infty}(-\beta)^{n}{\mathfrak K}_{1,n}.
\end{equation}
Let us define 
\begin{align}\label{Gamma_m}
\Gamma_{m}\equiv\overset{\infty}{\underset{n=m}\sum}(-\beta)^n{\rm K}_{1,m}^{(n)}
\end{align}
 as a summation of powers series of $\beta$. The expression of ${\rm K}_{1,m}^{(n)}$ is given by eq.(\ref{K1mn}). We point out that only connected open sub-chains contribute to the functions $\Gamma_m$. We define the function $\varphi(\lambda)$ in terms of the functions $\Gamma_m$ as follows 
\begin{align}\label{defvphi}
\varphi(\lambda)=\overset{\infty}{\underset{m=1}{\sum}}\frac{\Gamma_{m}}{\lambda^m},
\end{align}
 where $\lambda$ is a parameter. In \ref{apdC} we show that the 
\begin{equation}\label{s(lamd)}
\xi=\sum_{n=0}^{\infty}\frac{{\rm d}^n}{{\rm d}\lambda^n}\left(\frac{\varphi(\lambda)^{n+1}}{(n+1)!}\right)\bigg|_{\lambda=1}.
\end{equation}

From eqs.(\ref{k_rn}) and (\ref{Wbt}) we have that the connected and disconnected open sub-chains contribute to the grand potential per site. In the thermodynamic limit we have from eqs.(\ref{Gamma_m})-(\ref{s(lamd)}) that all these sub-chains can be derived only from a sum of connected open sub-chains (for details see \ref{apdC}). We finally write the grand potential per site simply as
\begin{align}\label{WTheta1}
{\cal W}(\beta,\mu)&=-\frac{1}{\beta}\left\{\ln({\rm tr}_1({\bf 1}))+\ln(1+\xi)\right\}.
\end{align}

Due to the fact that $\xi$ can be derived from function $\varphi(\lambda)$ one has the consequence that a much smaller number of terms has to be calculated at each $\beta$-order in $\beta$-expansion of ${\cal W}(\beta,\mu)$.  The weight of each g-trace in the eq.(\ref{WTheta1}) is already included in the definition of $\xi$.

\section{The Heisenberg XXZ Model}

Let us consider the Hamiltonian of the well known anisotropic one-dimensional Heisenberg XXZ model with spin-$\frac{1}{2}$\cite{gaudin,luther}
\begin{equation}\label{classicXXZ}
{\mathbb H} = \frac{1}{2}\sum_{j=1}^{N}(\s^{x}_{j}\s^{x}_{j+1}+\s^{y}_{j}\s^{y}_{j+1}+\Delta \s^{z}_{j}\s^{z}_{j+1}-2h \s^{z}_{j}),
\end{equation}
where $h$ is the external magnetic field  and $(\s^{x}_j,\s^{y}_j,\s^{z}_j)$ are the Pauli matrices at the $j$th site on a periodic chain with $N$ space sites and $\Delta$ is called the anisotropy parameter.  For  $\Delta>0$ $(\Delta<0)$ we have a repulsive (attractive) interaction core. The case $\Delta=1$ $(\Delta=-1)$ corresponds to the isotropic antiferromangetic (ferromagnetic) Heisenberg model with fully polarized ground state.

Through the Jordan-Wigner transformation\cite{Jordan}, the Hamiltonian (\ref{classicXXZ}) is mapped on a spinless fermionic model, whose Hamiltonian is 
\begin{equation}\label{XXZ}
{\mathbb H}=\sum_{j=1}^{N}\big(t({\pmb a}_j^{\dag}{\pmb a}_{j+1}+{\pmb a}_{j+1}^{\dag}{\pmb a}_{j})+V{\pmb n}_j{\pmb n}_{j+1}+E{\pmb n}_j\big)+N(h+\frac{\Delta}{2}),
\end{equation}
 with $V=2\Delta$, $E=-2h-2\Delta$, where ${\pmb n}_i={\pmb a}_i^{\dag}{\pmb a}_i$ and  ${\pmb a}_i^{\dag}({\pmb a}_i)$ is the fermionic creation(annihilation) operator at site $j$. The operators ${\pmb a}_i^{\dag}$ and ${\pmb a}_i$ satisfy the usual anti-commutation relations of the fermionic fields.  The hopping constant $t$ is included in Hamiltonian (\ref{XXZ}) only to help us counting the powers of the terms that contribute to a given order $\beta^n$ in the $\beta$-expansion of the Helmholtz free energy per site. Throughout our calculations, we take $t=1$. The term $N(h+\frac{\Delta}{2})$ leads to a shift in the energy of the ground state of the model.

\subsection{High Temperature Expansion}
In this section we calculate the $\beta$-expansion of the Helmholtz free energy per site of the XXZ model, whose Hamiltonian is given by eq.(\ref{XXZ}), up to order $\beta^3$.

In order to apply the method derived in section 2, we write Hamiltonian\eqref{classicXXZ} as: ${\mathbb H}=\sum_{i=1}^{N}{\bf H}_{i,i+1}$, where
\begin{equation}\label{subXXZ}
{\bf H}_{i,i+1}\equiv{\bf E}_{i,i+1}+{\bf T}^{+}_{i,i+1}+{\bf T}^{-}_{i,i+1}+{\bf V}_{i,i+1},
\end{equation}
and each term on the r.h.s. of eq.\eqref{subXXZ} is defined as: ${\bf E}_{i,i+1}\equiv\frac{1}{2}E({\pmb n}_i+{\pmb n}_{i+1})$, ${\bf T}^{-}_{i,i+1}\equiv t{\pmb a}_{i}^{\dag}{\pmb a}_{i+1}$, ${\bf T}^{+}_{i,i+1}\equiv t{\pmb a}_{i+1}^{\dag}{\pmb a}_{i}$ and ${\bf V}_{i,i+1}\equiv V{\pmb n}_i{\pmb n}_{i+1}$.

Our aim is to obtain the Helmholtz free energy per site of the XXZ model in the thermodynamic limit. For doing that we need the auxiliary function $\varphi(\lambda)$ (see eq.\eqref{defvphi}) of the XXZ model. This function is calculated from the functions $\Gamma_m$ that are written in terms of ${\rm K}_{1,m}^{(n)}$ (see eq.\eqref{Gamma_m}).
Since we intend to calculate the $\beta$-expansion  of the Helmholtz free energy per site of the XXZ model up to order $\beta^3$, we need to obtain the analytical expressions of $\Gamma_1,\dots,\Gamma_4$, namely,
\begin{subequations}\label{Gamma14}
\begin{align}
\Gamma_1&=-\beta{\rm K}_{1,1}^{(1)}+\beta^2{\rm K}_{1,1}^{(2)}-\beta^3{\rm K}_{1,1}^{(3)}+\beta^4{\rm K}_{1,1}^{(4)}+{\cal O}(\beta^5),\\
\Gamma_2&=\beta^2{\rm K}_{1,2}^{(2)}-\beta^3{\rm K}_{1,2}^{(3)}+\beta^4{\rm K}_{1,2}^{(4)}+{\cal O}(\beta^5),\\
\Gamma_3&=-\beta^3{\rm K}_{1,3}^{(3)}+\beta^4{\rm K}_{1,3}^{(4)}+{\cal O}(\beta^5),\\
\Gamma_4&=\beta^4{\rm K}_{1,4}^{(4)}+{\cal O}(\beta^5).
\end{align}
\end{subequations}
In appendix D the functions ${\rm K}_{1,m}^{(n)}$ in eqs.\eqref{Gamma14} are written as sums of normalized traces.

In particular, for Hamiltonian \eqref{XXZ} the hopping terms (${\bf T}^+_{i,i+1}$ and ${\bf T}^-_{i,i+1}$) in the normalized traces only contribute to those functions ${\rm K}_{1,m}^{(n)}$ where we have ${\bf H}_{i,i+1}^{l}$ with $l>1$.  Since we are calculating traces, the number of ${\bf T}^+_{i,i+1}$'s in a given normalized trace has to be equal to the number of terms ${\bf T}^-_{i,i+1}$'s in it. Otherwise, the normalized trace is null.

Calculating explicitly the normalized traces that contribute to ${\rm K}_{1,m}^{(n)}$ (see appendix D) for the XXZ model, we obtain
\begin{itemize}
\begin{subequations}
\item $m=1$
\begin{align}
{\rm K}_{1,1}^{(1)}&=\frac{E}{2}+\frac{V}{4},\\
{\rm K}_{1,1}^{(2)}&=\frac{t^2}{4}+\frac{3E^2}{16}+\frac{EV}{4}+\frac{V^2}{8},\\
{\rm K}_{1,1}^{(3)}&=\frac{5E^3}{96}+\frac{V^3}{24}+\frac{Et^2}{8}+\frac{VE^2}{8}+\frac{EV^2}{8},\\
{\rm K}_{1,1}^{(4)}&=\frac{t^4}{48}+\frac{E^2t^2}{32}+\frac{3E^4}{256}+\frac{VE^3}{24}+\frac{E^2V^2}{16}+\frac{EV^3}{24}+\frac{V^4}{96}.
\end{align}
\end{subequations}
\item $m=2$
\begin{subequations}
\begin{align}
{\rm K}_{1,2}^{(2)}&=\frac{5E^2}{16}+\frac{3EV}{8}+\frac{V^2}{8},\\
{\rm K}_{1,2}^{(3)}&=\frac{E^3}{4}+\frac{V^3}{8}+\frac{Et^2}{4}+\frac{17VE^2}{32}+\frac{7EV^2}{16}+\frac{Vt^2}{8},\\
{\rm K}_{1,2}^{(4)}&=\frac{t^4}{24}+\frac{5E^2t^2}{24}+\frac{95E^4}{768}+\frac{25VE^3}{64}+\frac{33E^2V^2}{64}+\frac{5EV^3}{16}+\frac{7V^4}{96}+\frac{t^2V^2}{24}+\frac{EVt^2}{6}.
\end{align}
\end{subequations}
\item $m=3$
\begin{subequations}
\begin{align}
{\rm K}_{1,3}^{(3)}&=\frac{3E^3}{16}+\frac{V^3}{16}+\frac{23VE^2}{64}+\frac{EV^2}{4},\\
{\rm K}_{1,3}^{(4)}&=\frac{41E^2t^2}{192}+\frac{59E^4}{256}+\frac{85VE^3}{128}+\frac{101E^2V^2}{128}+\frac{7EV^3}{16}+\frac{7t^2V^2}{96}+\frac{23EVt^2}{96}+\frac{3V^4}{32}.
\end{align}
\end{subequations}
\item $m=4$
\begin{subequations}
\begin{align}
{\rm K}_{1,4}^{(4)}&=\frac{29E^4}{256}+\frac{19VE^3}{64}+\frac{5E^2V^2}{16}+\frac{5EV^3}{32}+\frac{V^4}{32}.
\end{align}
\end{subequations}
\end{itemize}

Substituting those expressions in eqs.\eqref{Gamma14} (taking $t=1$) and those in eq.\eqref{defvphi} we obtain the auxiliary function $\varphi(\lambda)$ for the XXZ model. Using the algebraic language {\sf MAPLE 5.1} we calculate the Helmholtz free energy per site of the model up to order $\beta^3$, that is,

\begin{align}\label{wb4}
{\cal W}(\beta)=\frac{\ln(2)}{\beta}+{\cal W}_I(\beta)+{\cal W}_F(\beta)-\frac{\Delta}{8}\beta^2+\frac{1}{4}\Big(h^2+\frac{\Delta^2}{6}\Big)\beta^3+{\cal O}(\beta^4),
\end{align}
where ${\cal W}_{I}(\beta)$ is the Helmholtz free energy per site of the Ising model\cite{L_baxter,bjp} up to order $\beta^3$, that is,
\begin{equation}
{\cal W}_{I}(\beta)=-\frac{\ln(2)}{\beta}-\Big(\frac{h^2}{2}+\frac{\Delta^2}{8}\Big)\beta+\frac{h^2\Delta}{2}\beta^2+\Big(\frac{h^4}{12}-\frac{h^2\Delta^2}{4}+\frac{\Delta^4}{192}\Big)\beta^3+{\cal O}(\beta^4)
\end{equation}
and ${\cal W}_F(\beta)$ is the Helmholtz free energy per site of the free fermion model\cite{bjp,klumper93} up to same $\beta$-order, 
\begin{equation}
{\cal W}_{F}(\beta)=-\frac{\ln(2)}{\beta}-\frac{t^2}{4}\beta-\frac{t^4}{32}\beta^3+{\cal O}(\beta^4).
\end{equation}

We rewrite the Helmholtz free energy per site (eq.\eqref{wb4}) using the constants of Hamiltonian in reference \cite{destri}. Comparing our results with eq.(6.5) of this reference, we conclude: {\it i}) both results agree up to order $\beta$; {\it ii}) our coefficient at order $\beta^2$ is: $-J^3\cos(\gamma)+J\cos(\gamma)h^2$ which means that we have two misprints in eq.(6.5) of reference \cite{destri}, that is: the sign of $-J^3\cos(\gamma)\beta^2$ and is missing a power of $J$ multiplying $h^2$ (the two terms that contribute to order $\beta^2$ in eq.(6.5) have different dimensions); {\it iii}) it is missing the term $-J^2\cos^2(\gamma)h^2\beta^3$ in eq.(6.5) of reference \cite{destri} (we point out that without this term we do not recover the limiting case of Ising model\cite{L_baxter,bjp} ($t=0$) from XXZ model), morever the other terms at order $\beta^3$ in eq.(6.5) have the following misprints: $-\frac{1}{4}J^4\beta^3$ should be $\frac{1}{2}J^4\beta^3$ and  the term $-\frac{1}{6}J^4\cos^2(\gamma)\beta^3$ should be $\frac{2}{3}J^4\cos^2(\gamma)\beta^3$.  With the previous corrections we get the Helmholtz free energy per site of the Ising model and the free fermion model as limiting cases of the XXZ model.

In reference \cite{destri} Destri and de Vega obtained the Helmholtz free energy per site of the XXZ model as a non-linear integral equation (NLIE), as similarly obtained by Kl\"umper\cite{klumper93}. Once our disagreement with the result of reference\cite{destri} is proportional $h^2$ at order $\beta^3$, we calculate the $\beta$-expansion of the magnetic susceptibility $\chi(\beta)$ ($\chi(\beta)=-\frac{\partial^2{\mathcal W}(\beta)}{\partial h^2}$), obtained from eq.\eqref{wb4}. In figure 1 we plot the curve of the magnetic susceptibility derived from NLIE in reference \cite{klumper93,destri}, the $\beta$-expansion of $\chi(\beta)$ derived in reference\cite{destri} and our own $\beta$-expansion of $\chi(\beta)$. From figure 1 we see that our result with $\Delta=\frac{1}{2}$ and $\Delta=\frac{\sqrt{2}}{2}$ coincides with the one derived from the NLIE in reference\cite{klumper93,destri} for $\beta$ up to $0.3$, while de $\beta$-expansion of Destri and de Vega does not.
\begin{center}
  \psfrag{Dh0}[]{Magnetic Susceptibility}
  \psfrag{Dh1}[]{Magnetic Susceptibility}
  \psfrag{b}[]{$\beta$}
  \psfrag{X}[]{$\chi(\beta)$}

  \psfrag{0.1}{\small$0.1$}
  \psfrag{0.2}{\small$0.2$}
  \psfrag{0.3}{\small$0.3$}
  \psfrag{0.4}{\small$0.4$}

  \psfrag{0}{\small$0$}
  \psfrag{0.25}{\small$0.25$}
  \psfrag{0.5}{\small$0.5$}
  \psfrag{0.75}{\small$0.75$}
  \psfrag{1}{\small$1$}
  \psfrag{f1}{$\Delta=\frac{1}{2}$}
  \psfrag{f2}{$\Delta=\frac{\sqrt{2}}{2}$}
\includegraphics[width=14cm]{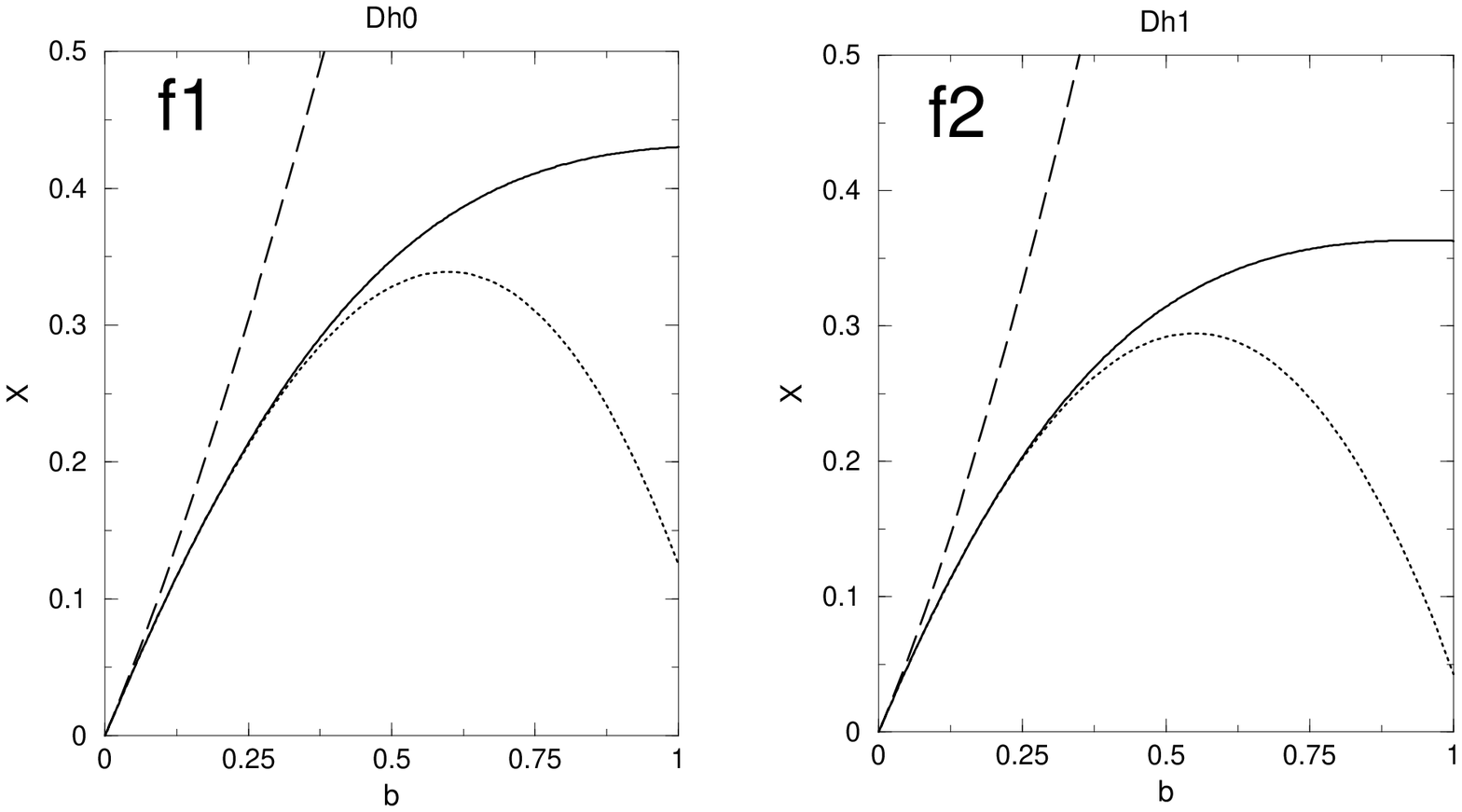}
\begin{figure}[h]
\caption[fig1]{ The pointed line corresponds the $\beta$-expansion (\ref{wb4}), the dashed line represents the $\beta$-expansion obtained by Destri e de Vega\cite{destri} and the solid line corresponds  the numerical solution of the NLIE in reference\cite{klumper93,destri}.}
\end{figure}
\end{center}

As a final check of our $\beta$-expansion\eqref{wb4} we plot in figure 2 the third partial derivative with respect to $\beta$ of the function $\beta{\mathcal W}(\beta)$. We compare the numerical solution for this function obtained from the NLIE\cite{klumper93,destri}, in the same reference, and our results derived from eq.\eqref{wb4}. In the region $\beta \approx 0$, the function $\frac{\partial^3}{\partial\beta^3}\big(\beta{\mathcal W}(\beta)\big)$ is a straight line. From figure 2 we see that  the result derived from eq.\eqref{wb4} is identical to the one from the NLIE for $\beta$ up to 0.1.  For $h\ne 0$ in figure 2 we get that the function $\frac{\partial^3}{\partial\beta^3}\big(\beta{\mathcal W}(\beta)\big)$ derived from the $\beta$-expansion in reference does not even touch the exact curve.
\begin{center}
  \psfrag{Dh0}[]{$h=0$}
  \psfrag{Dh1}[]{$h=0.25$}
  \psfrag{b}[]{$\beta$}
  \psfrag{X}[]{$\frac{\partial^3}{\partial\beta^3}\big(\beta{\mathcal W}(\beta)\big)$}
  \psfrag{0.1}{\small$0.1$}
  \psfrag{0.2}{\small$0.2$}
  \psfrag{0.3}{\small$0.3$}
  \psfrag{0}{\small$0$}
  \psfrag{-0.1}{\small$-0.1$}
  \psfrag{-0.2}{\small$-0.2$}
  \psfrag{-0.3}{\small$-0.3$}
  \psfrag{-0.4}{\small$-0.4$}
  \psfrag{-0.6}{\small$-0.6$}
  \psfrag{-0.8}{\small$-0.8$}
  \psfrag{f1}{$\Delta=\frac{1}{2}$}
  \psfrag{f2}{$\Delta=\frac{1}{2}$}
\includegraphics[width=14cm]{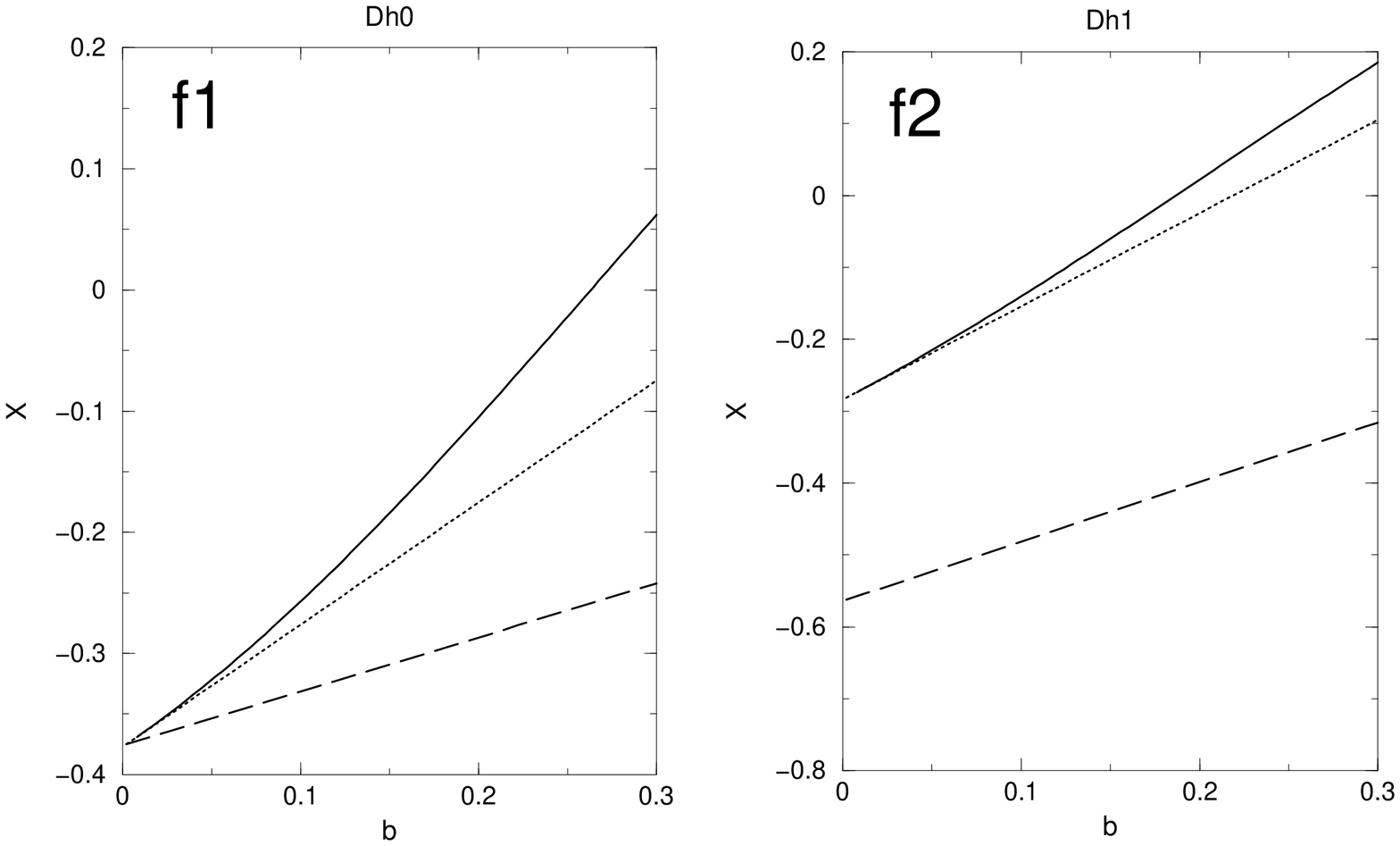}

\begin{figure}[h]
\caption[fig1]{ We plot the function $\frac{\partial^3}{\partial\beta^3}\big(\beta{\mathcal W}(\beta)\big)$ versus $\beta$. The pointed line (a straight line) corresponds to the results obtained from (\ref{wb4}), the dashed line  the $\beta$-expansion obtained by Destri e de Vega\cite{destri} and the solid line the numerical solution of the NLIE\cite{klumper93,destri}.}
\end{figure}
\end{center}

\section{Conclusions}

There are several interesting one-dimensional models with translational invariance, with interactions between first neighbors only, satisfying periodic boundary conditions. They are generically called chain models, some which having the property of integrability and to which the Bethe ansatz has been applied to solve them  exactly at $T=0$ and at finite $T$. Their thermodynamical properties are obtained through coupled integral equations which demand numerical analysis to be solved (as examples see references \cite{takahashi_sk,kuniba,bariev,ramos,juttner}).

In this paper we present a new analytical method to get the $\beta$-expansion coefficients of the grand potential per site, for translationally invariant chain models with nearest-neighbors interactions.  Differently  from our previous works\cite{irazietjmp,irazietphya,iraziet_sub} we calculate directly the $\beta$-expansion of the grand potential per site ${\cal W}(\beta,\mu)$. The weight of each sub-chain in ${\cal W}(\beta,\mu)$ is obtained, and we show explicitly that the grand potential per site is an intensive quantity.  In the thermodynamic limit $(N\rightarrow\infty)$ we show that ${\cal W}(\beta,\mu)$ can be derived from an auxiliary function $\varphi(\lambda)$ that is written only in terms of open connected sub-chains.  The present approach  gives a $\beta$-expansion of ${\cal W}(\beta,\mu)$, whose coefficient of order $\beta^n$ can be obtained exactly for arbitrary value of $n$. The existence of this auxiliary function allow us to get higher order terms in the $\beta$-expansion of ${\cal W}(\beta,\mu)$ than we were able before, in references \cite{irazietjmp,irazietphya,iraziet_sub}. The coefficient of order $\beta^n$ is analytically obtained, and it is exact. The number of terms (traces) to be calculated when derivig each coefficient in $\beta$-expansion of ${\cal W}(\beta,\mu)$ is much smaller than before.

The present method applies to chains in both real space and momentum space, and the result given by eq.(\ref{WTheta1}) is valid for both classic and quantum models (bosonic and/or fermionic fields).

As an illustrative application of the method, we consider the spin-$\frac{1}{2}$ Heisenberg XXZ chain model and derive its Helmholtz free energy per site up to order $\beta^3$. Our $\beta$-expansion \eqref{wb4} recovers the Helmholtz free energy per site in two limiting cases: the Ising model ($t=0$) and the free fermion model ($\Delta=h=0$). We compare our high temperature expansion of the ${\mathcal W}(\beta)$ eq.\eqref{wb4} of the XXZ model with the one derived by Destri and de Vega\cite{destri}.  We show that this last one has a misprint in eq.(6.5). 

As a final word we want to point out that the present approach can be well applied to models that do not have the property of integrability.  Besides, once we have closed expressions for the coefficients of the $\beta$-expansion of ${\mathcal W}(\beta)$ the method can be implemented through a procedure in an algebraic language to get high $\beta$-order terms of the high temperature expansion of ${\mathcal W}(\beta)$ for the model under interest.  We hope to report in the future progress in this direction.

\section*{ Acknowledgments}

S. M. de S. is in debt with L. M. Chaves for very useful discussions. The authors are in debt we E. V. Corr\^ea Silva for reading the manuscript. O. R. thanks CAPES for financial support. M.T.T and S.M. de S. thank CNPq for partial financial support and S.M. de S. thank FAPEMIG for partial financial support. M.T.T. also thanks FAPERJ for partial financial support.

\begin{appendix}
\setcounter         {section}{0}
\setcounter         {equation}{0}
\def\theequation{A.\arabic{equation}}
\def\thesection{Appendix A}
\section{Graphic representation of normalized traces}
We have a better visualization of the normalized traces that contribute to $\bra{\mathbb K}^n\ket$ through a graphic representation. In the graph, the operator ${\bf K}_{i,i+1}$ is represented by a straight line that links the sites $i$ and $i+1$. In the case we have a product of operators ${\bf K}_{i,i+1}$'s for each operator a line is drawn at a different step of the ladder.  It is important in the normalized trace the order in which the operators ${\bf K}_{i,i+1}$ appear in the string. We take this order into account in our graph representation of the normalized traces. To exemplify how to take into account the order of the operators in the string, we consider the following examples:
\begin{eqnarray}\label{trv b2}
\bra{\bf K}_{i,i+1}\ket
\quad\longrightarrow\quad
\begin{picture}(100,10)
\thicklines
\put(40,2){\line(1,0){30}}
\multiput(0,2)(4,0){30}{\line(1,0){1}}
\multiput(10,2)(30,0){4}{\circle{1}}
\multiput(10,2)(30,0){4}{\circle{2}}
\put(0,-8){\footnotesize{$i-1$}}
\put(38,-8){\footnotesize{$i$}}
\put(62,-8){\footnotesize{$i+1$}}
\put(92,-8){\footnotesize{$i+2$}}
\end{picture}
\end{eqnarray}

\begin{eqnarray}
\bra{\bf K}_{i,i+1}{\bf K}_{i+1,i+2}\ket
\quad\longrightarrow\quad
\begin{picture}(100,20)
\thicklines
\put(10,2){\line(1,0){30}}
\put(40,12){\line(1,0){30}}
\multiput(40,12)(30,0){2}{\circle{1}}
\multiput(40,12)(30,0){2}{\circle{2}}
\multiput(0,2)(4,0){30}{\line(1,0){1}}
\multiput(40,2)(0,2){5}{\line(0,1){1}}
\multiput(70,2)(0,2){5}{\line(0,1){1}}
\multiput(10,2)(30,0){4}{\circle{1}}
\multiput(10,2)(30,0){4}{\circle{2}}
\put(10,-8){\footnotesize{$i$}}
\put(30,-8){\footnotesize{$i+1$}}
\put(62,-8){\footnotesize{$i+2$}}
\put(92,-8){\footnotesize{$i+3$}}
\end{picture}
\end{eqnarray}

\begin{eqnarray}\label{trv a}
\bra{\bf K}_{i,i+1}{\bf K}_{i-1,i}{\bf K}_{i+1,i+2}\ket
\quad\longrightarrow\quad
\begin{picture}(110,20)
\thicklines
\put(40,2){\line(1,0){30}}
\put(10,12){\line(1,0){30}}
\put(70,24){\line(1,0){30}}
\multiput(10,12)(30,0){2}{\circle{1}}
\multiput(10,12)(30,0){2}{\circle{2}}
\multiput(70,24)(30,0){2}{\circle{1}}
\multiput(70,24)(30,0){2}{\circle{2}}
\multiput(0,2)(4,0){30}{\line(1,0){1}}
\multiput(10,2)(0,2){5}{\line(0,1){1}}
\multiput(40,2)(0,2){5}{\line(0,1){1}}
\multiput(70,2)(0,2){11}{\line(0,1){1}}
\multiput(100,2)(0,2){11}{\line(0,1){1}}
\multiput(10,2)(30,0){4}{\circle{1}}
\multiput(10,2)(30,0){4}{\circle{2}}
\put(0,-8){\footnotesize{$i-1$}}
\put(38,-8){\footnotesize{$i$}}
\put(62,-8){\footnotesize{$i+1$}}
\put(92,-8){\footnotesize{$i+2$}}
\end{picture}
\end{eqnarray}

\begin{eqnarray}\label{trv b4}
\bra{\bf K}_{i,i+1}{\bf K}_{i+1,i+2}{\bf K}_{i-1,i}\ket
\quad\longrightarrow\quad
\begin{picture}(110,20)
\thicklines
\put(40,2){\line(1,0){30}}
\put(10,24){\line(1,0){30}}
\put(70,12){\line(1,0){30}}
\multiput(10,24)(30,0){2}{\circle{1}}
\multiput(10,24)(30,0){2}{\circle{2}}
\multiput(70,12)(30,0){2}{\circle{1}}
\multiput(70,12)(30,0){2}{\circle{2}}
\multiput(0,2)(4,0){30}{\line(1,0){1}}
\multiput(10,2)(0,2){11}{\line(0,1){1}}
\multiput(40,2)(0,2){11}{\line(0,1){1}}
\multiput(70,2)(0,2){5}{\line(0,1){1}}
\multiput(100,2)(0,2){5}{\line(0,1){1}}
\multiput(10,2)(30,0){4}{\circle{1}}
\multiput(10,2)(30,0){4}{\circle{2}}
\put(0,-8){\footnotesize{$i-1$}}
\put(38,-8){\footnotesize{$i$}}
\put(62,-8){\footnotesize{$i+1$}}
\put(92,-8){\footnotesize{$i+2$}}
\end{picture}
\end{eqnarray}
The graphs (\ref{trv b2})-(\ref{trv b4}) are {\it connected} graphs, which means that the normalized trace can not be written as a product of normalized traces of smaller number of operators ${\bf K}_{i,i+1}$.

\setcounter         {equation}{0}
\def\theequation{B.\arabic{equation}}
\def\thesection{Appendix B}
\section{Calculation of $\bra\Kp^n\ket$}\label{apdB}

In this appendix we give the explicit definition of ${\rm K}_{r,m}^{(4)}$ in terms of which we write $\bra\Kp^4\ket$ .  The expression of $\bra\Kp^4\ket$ plus eqs.(\ref{K 1 2 3}) will allow us to write down $\bra\Kp^n\ket$ for arbitrary $n$.

 We define ${\rm K}_{1,m}^{(4)}$ as
\begin{subequations}
\begin{align}
{\rm K}_{1,1}^{(4)}&=\frac{\bra{\bf K}_{1,2}^4\ketg}{4!},\label{M114}\\ {\rm K}_{1,2}^{(4)}&=\frac{\bra{\bf K}_{1,2}^3{\bf K}_{2,3}\ketg}{3!}+\frac{\bra{\bf K}_{1,2}{\bf K}_{2,3}^3\ketg}{3!}+\frac{\bra{\bf K}_{1,2}^2{\bf K}_{2,3}^2\ketg}{2!\;2!},\label{M124}\\ {\rm K}_{1,3}^{(4)}&=\frac{\bra{\bf K}_{1,2}^2{\bf K}_{2,3}{\bf K}_{3,4}\ketg}{2!}+\frac{\bra{\bf K}_{1,2}{\bf K}_{2,3}^2{\bf K}_{3,4}\ketg}{2!}+\frac{\bra{\bf K}_{1,2}{\bf K}_{2,3}{\bf K}_{3,4}^2\ketg}{2!},\label{M134}\\ {\rm K}_{1,4}^{(4)}&=\bra{\bf K}_{1,2}{\bf K}_{2,3}{\bf K}_{3,4}{\bf K}_{4,5}\ketg .\label{M144}
\end{align}
\end{subequations}

The other traces can be written as products of traces of lower order and they are
\begin{subequations}
\begin{align}
{\rm K}_{2,2}^{(4)} =&\frac{\bra{\bf K}_{1,2}^3{\bf K}_{3,4}\ketg}{3!}+\frac{\bra{\bf K}_{1,2}{\bf K}_{3,4}^3\ketg}{3!}+\frac{\bra{\bf K}_{1,2}^2{\bf K}_{3,4}^2\ketg}{2!\;2!},\label{M224}\\ {\rm K}_{2,3}^{(4)} =&\frac{\bra{\bf K}_{1,2}^2{\bf K}_{2,3}{\bf K}_{4,5}\ketg}{2!}+\frac{\bra{\bf K}_{1,2}{\bf K}_{2,3}^2{\bf K}_{4,5}\ketg}{2!}+\frac{\bra{\bf K}_{1,2}{\bf K}_{2,3}{\bf K}_{4,5}^2\ketg}{2!}+\label{M234}\nonumber\\&+\frac{\bra{\bf K}_{1,2}^2{\bf K}_{3,4}{\bf K}_{4,5}\ketg}{2!}+\frac{\bra{\bf K}_{1,2}{\bf K}_{3,4}^2{\bf K}_{4,5}\ketg}{2!}+\frac{\bra{\bf K}_{1,2}{\bf K}_{3,4}{\bf K}_{4,5}^2\ketg}{2!},
\\ {\rm K}_{2,4}^{(4)}=&\bra{\bf K}_{1,2}{\bf K}_{2,3}{\bf K}_{3,4}{\bf K}_{5,6}\ketg+\bra{\bf K}_{1,2}{\bf K}_{3,4}{\bf K}_{4,5}{\bf K}_{5,6}\ketg+\bra{\bf K}_{1,2}{\bf K}_{2,3}{\bf K}_{4,5}{\bf K}_{5,6}\ketg,\label{M244}\\ {\rm K}_{3,3}^{(4)} =&\frac{\bra{\bf K}_{1,2}^2{\bf K}_{3,4}{\bf K}_{5,6}\ketg}{2!}+\frac{\bra{\bf K}_{1,2}{\bf K}_{3,4}^2{\bf K}_{5,6}\ketg}{2!}+\frac{\bra{\bf K}_{1,2}{\bf K}_{3,4}{\bf K}_{5,6}^2\ketg}{2!},\label{M334}\\ {\rm K}_{3,4}^{(4)} =&\bra{\bf K}_{1,2}{\bf K}_{2,3}{\bf K}_{4,5}{\bf K}_{6,7}\ketg+\bra{\bf K}_{1,2}{\bf K}_{2,3}{\bf K}_{4,5}{\bf K}_{6,7}\ketg+\bra{\bf K}_{1,2}{\bf K}_{2,3}{\bf K}_{4,5}{\bf K}_{6,7}\ketg,\label{M344}\\ {\rm K}_{4,4}^{(4)} =&\bra{\bf K}_{1,2}{\bf K}_{3,4}{\bf K}_{5,6}{\bf K}_{7,8}\ketg\label{M444}.
\end{align}
\end{subequations}
From those definitions, the normalized trace $\bra\Kp^4\ket$ is written as
\begin{align}\label{K1_4}
\frac{\bra\Kp^4\ket}{4!}&= N{\rm K}_{1,1}^{(4)}+N{\rm K}_{1,2}^{(4)}+N{\rm K}_{1,3}^{(4)}+N{\rm K}_{1,4}^{(4)}+\frac{N(N-3)}{2!}{\rm K}_{2,2}^{(4)}+\frac{N(N-4)}{2!}{\rm K}_{2,3}^{(4)}+\frac{N(N-5)}{2!}{\rm K}_{2,4}^{(4)}\nonumber\\ &\quad+\frac{N(N-4)(N-5)}{3!}{\rm K}_{3,3}^{(4)}+\frac{N(N-5)(N-6)}{3!}{\rm K}_{3,4}^{(4)}+\frac{N(N-5)(N-6)(N-7)}{4!}{\rm K}_{4,4}^{(4)}\;.
\end{align}
We use the following notations to write the coefficients of eqs.(\ref{K 1 2 3}) and eq.(\ref{K1_4}):

\begin{align}\label{c11_4}
{\rm C}^N_{1,1}&=N,&&&&\nonumber\\ {\rm C}^N_{1,2}&=N,&{\rm C}^N_{2,2}&=\frac{N(N-3)}{2!},&&\nonumber\\ {\rm C}^N_{1,3}&=N,&{\rm C}^N_{2,3}&=\frac{N(N-4)}{2!},&{\rm C}^N_{3,3}&=\frac{N(N-4)(N-5)}{3!}&\nonumber\\ {\rm C}^N_{1,4}&=N,&{\rm C}^N_{2,4}&=\frac{N(N-5)}{2!},&{\rm C}^N_{3,4}&=\frac{N(N-5)(N-6)}{3!},&{\rm C}^N_{4,4}&=\frac{N(N-5)(N-6)(N-7)}{4!}.
\end{align}

Replacing the definitions given by eqs.(\ref{c11_4}) in eqs.(\ref{K 1 2 3}) and (\ref{K1_4}), we obtain a compact way of writing the normalized traces
\begin{subequations}
\begin{align}
\bra\Kp\ket&= {\rm C}^N_{1,1}{\rm K}_{1,1}^{(1)}\;,\\ \frac{\bra\Kp^2\ket}{2!}&= {\rm C}^N_{1,1}{\rm K}_{1,1}^{(2)}+{\rm C}^N_{1,2}{\rm K}_{1,2}^{(2)}+{\rm C}^N_{2,2}{\rm K}_{2,2}^{(2)}\;,\\ \frac{\bra\Kp^3\ket}{3!}&= {\rm C}^N_{1,1}{\rm K}_{1,1}^{(3)}+{\rm C}^N_{1,2}{\rm K}_{1,2}^{(3)}+{\rm C}^N_{1,3}{\rm K}_{1,3}^{(3)}+{\rm C}^N_{2,2}{\rm K}_{2,2}^{(3)}+{\rm C}^N_{2,3}{\rm K}_{2,3}^{(3)}+{\rm C}^N_{3,3}{\rm K}_{3,3}^{(3)}\;,\\ \frac{\bra\Kp^4\ket}{4!}&= {\rm C}^N_{1,1}{\rm K}_{1,1}^{(4)}+{\rm C}^N_{1,2}{\rm K}_{1,2}^{(4)}+{\rm C}^N_{1,3}{\rm K}_{1,3}^{(4)}+{\rm C}^N_{1,4}{\rm K}_{1,4}^{(4)}+{\rm C}^N_{2,2}{\rm K}_{2,2}^{(4)}+{\rm C}^N_{2,3}{\rm K}_{2,3}^{(4)}+{\rm C}^N_{2,4}{\rm K}_{2,4}^{(4)}+\nonumber\\&\quad +{\rm C}^N_{3,3}{\rm K}_{3,3}^{(4)}+{\rm C}^N_{3,4}{\rm K}_{3,4}^{(4)}+{\rm C}^N_{4,4}{\rm K}_{4,4}^{(4)}\;.
\end{align}
\end{subequations}
From eqs.(\ref{c11_4}) we see that coefficient that multiply ${\rm K}_{r,m}^{(n)}$ is independent of $n$ and it is equal to
\begin{subequations}\label{c1m}
\begin{align}
{\rm C}^N_{1,m}&=N\;,\hskip1cm 1\leqslant m,\\ {\rm C}^N_{2,m}&=\frac{N(N-m-1)}{2!}\;,\hskip1cm 2\leqslant m,\\ {\rm C}^N_{3,m}&=\frac{N(N-m-1)(N-m-2)}{3!}\;,\hskip1cm 3\leqslant m, \\ {\rm C}^N_{4,m}&=\frac{N(N-m-1)(N-m-2)(N-m-3)}{4!}\;,\hskip1cm 4\leqslant m.
\end{align}
\end{subequations}
Afterwards for any $r\leqslant m\leqslant {\rm min}(n,N)$, we obtain a general relation from eqs.(\ref{c1m}) given by
\begin{equation}\label{crm}
{\rm C}^N_{r,m}=\frac{N}{r}\binom{N-m-1}{r-1}=\frac{1}{r}\sum_{k=1}^{r}(-1)^{r+k}k\binom{m+r-k-1}{r-k}\binom{N}{k}\;.
\end{equation}
such that 
\begin{equation}\label{TrnK Ap}
\frac{\bra{\Kp}^n\ket}{n!}=\sum_{r=1}^{[n,N]}\sum_{m=r}^{[n,N]}\frac{N}{r}\binom{N-m-1}{r-1} {\rm K}^{(n)}_{r,m}\;.
\end{equation}
We are using the notation: $[n,N]={\rm min}(n,N)$.
\setcounter         {equation}{0}
\def\theequation{C.\arabic{equation}}
\def\thesection{Appendix C}

\section{Relation between the auxiliary function $\varphi(\lambda)$ and ${\xi}$}\label{apdC}

In the eq.(\ref{Gamma_m}) we define the function $\Gamma_m$,
\begin{align}\label{Gamma_m_C}
\Gamma_{m}\equiv\overset{\infty}{\underset{n=m}\sum}(-\beta)^n{\rm K}_{1,m}^{(n)},
\end{align}
and the auxiliary function $\varphi(\lambda)$,
\begin{align}\label{varphi_gm_C}
\varphi(\lambda)\equiv\sum_{m=1}^{\infty}\frac{\Gamma_m}{\lambda^m}.
\end{align}
In this appendix we show that the grand potential per site, in the thermodynamic limit, can be obtained from derivatives of powers of the function $\varphi(\lambda)$.

 To calculate the grand potential per site (see eq.(\ref{Wbt})) we need to perform the sum
\begin{equation}\label{s_C}
\sum_{n=0}^{\infty}(-\beta)^{n}{\mathfrak K}_{1,n}\equiv \xi.
\end{equation}
Substituting the expression of ${\mathfrak K}_{1,n}$ (eq.(\ref{k_rn}) with $k=1$) in $\xi$ and taking into account that in ${\rm K}_{1,m}^{(n)}$ we have to have $n>m$, eq.(\ref{s_C}) becomes
\begin{align}
\xi=&\sum_{n=1}^{\infty}\sum_{m=1}^n(-\beta)^n{\rm K}_{1,m}^{(n)}-\frac{1}{2}\sum_{n=2}^{\infty}\sum_{m=2}^n(-\beta)^n\binom{m}{1}{\rm K}_{2,m}^{(n)}+\frac{1}{3}\sum_{n=3}^{\infty}\sum_{m=3}^n(-\beta)^n\binom{m+1}{2}{\rm K}_{3,m}^{(n)}+\dots\\
\equiv&\sum_{r=1}^{\infty}\varphi_r,
\end{align}
where $\varphi_r$ is such that
\begin{align}\label{varphi_r}
\varphi_r\equiv&\frac{(-1)^{r+1}}{r}\sum_{n=r}^{\infty}\sum_{m=r}^n(-\beta)^n\binom{m+r-2}{r-1}{\rm K}_{r,m}^{(n)}.
\end{align}
We drop the upper limit $[n,N]$ condition in the sum over $m$ (see eq.(\ref{k_rn})) because now we are taking first the thermodynamic limit $(N\rightarrow\infty)$, which means that we always have $n<N$.

Applying the equality
\begin{equation}
\sum_{n=r}^{\infty}\sum_{m=r}^{n}\dots=\sum_{m=r}^{\infty}\sum_{n=r}^{\infty}\dots
\end{equation}
on eq.(\ref{varphi_r}), we obtain 
\begin{align}
\varphi_r=\frac{(-1)^{r+1}}{r}\sum_{m=r}^{\infty}\binom{m+r-2}{r-1}\sum_{n=r}^{\infty}(-\beta)^n{\rm K}_{r,m}^{(n)}.
\end{align}
From eqs.(\ref{Gamma_m_C}) and (\ref{varphi_gm_C}) we have
\begin{align}\label{varphi}
\varphi(\lambda)=\sum_{m=1}^{\infty}\frac{1}{\lambda^m}\sum_{n=1}^{\infty}(-\beta)^n{\rm K}_{1,m}^{(n)}.
\end{align}
Our next step is to take powers of function $\varphi(\lambda)$. We begin with $\varphi(\lambda)^2$,
\begin{align}
\varphi(\lambda)^2=&\sum_{m_1,m_2=1}^{\infty}\frac{1}{\lambda^{m_1+m_2}}\sum_{n_1,n_2=1}^{\infty}(-\beta)^{n_1+n_2}{\rm K}_{1,m_1}^{(n_1)}{\rm K}_{1,m_2}^{(n_2)}\nonumber\\
=&\sum_{m=2}^{\infty}\frac{1}{\lambda^{m}}\sum_{n=2}^{\infty}(-\beta)^{n}\sump_{\{n_i\}}^n\sump_{\{m_i\}}^m{\rm K}_{1,m_1}^{(n_1)}{\rm K}_{1,m_2}^{(n_2)}\nonumber\\
=&\sum_{m=2}^{\infty}\frac{1}{\lambda^{m}}\sum_{n=2}^{\infty}(-\beta)^{n}{\rm K}_{2,m}^{(n)}.
\end{align}
For arbitrary $r$ we obtain
\begin{equation}
\varphi(\lambda)^r=\sum_{m=r}^{\infty}\frac{1}{\lambda^{m}}\sum_{n=r}^{\infty}(-\beta)^{n}{\rm K}_{r,m}^{(n)}
\end{equation}
and we note that
\begin{subequations}
\begin{align}
\varphi(\lambda)\big|_{\lambda=1}=\varphi_1\\
\frac{\rm d}{\rm d\lambda}\Big(\frac{\varphi(\lambda)^2}{2!}\Big)\Big|_{\lambda=1}=\varphi_2\\\nonumber
\vdots&\\
\frac{{\rm d}^{r-1}}{{\rm d}\lambda^{r-1}}\Big(\frac{\varphi(\lambda)^r}{r!}\Big)\Big|_{\lambda=1}=\varphi_r\;.
\end{align}
\end{subequations}
Finally we have
\begin{align}\label{def y}
\xi&=\sum_{r=1}^{\infty}\varphi_r\\\nonumber&=\sum_{r=1}^{\infty}\frac{{\rm d}^{r-1}}{{\rm d}\lambda^{r-1}}\Big(\frac{\varphi(\lambda)^r}{r!}\Big)\Big|_{\lambda=1},
\end{align}
which shows that the sum (\ref{s_C}) can be obtained from a function that contains only open connected sub-chains.

\setcounter         {equation}{0}
\def\theequation{D.\arabic{equation}}
\def\thesection{Appendix D}

\section{Useful relations between ${\rm K}_{1,m}^{(n)}$ and the normalized traces}\label{apdD}
In section 3 we calculate the $\beta$-expansion of the Helmholtz free energy per site of the XXZ model up to order $\beta^3$.  In order to do so, we need to write the functions ${\rm K}_{1,m}^{(n)}$ (see eq.\eqref{K1mn}) in terms of the normalized traces instead of the g-traces.

For any quantum system driven by a Hamiltonian with interactions  between first neighbors only, invariant under translations along the chain, and null chemical potential, we  have:

\begin{itemize}
\item $m=1$
\begin{subequations}
\begin{align}
{\rm K}_{1,1}^{(1)}&=\bra{\bf H}_{1,2}\ket,\\
{\rm K}_{1,1}^{(2)}&=\frac{1}{2!}\bra{\bf H}_{1,2}^2\ket,\\
{\rm K}_{1,1}^{(3)}&=\frac{1}{3!}\bra{\bf H}_{1,2}^3\ket,\\
{\rm K}_{1,1}^{(4)}&=\frac{1}{4!}\bra{\bf H}_{1,2}^4\ket.
\end{align}
\end{subequations}
\item $m=2$
\begin{subequations}
\begin{align}
{\rm K}_{1,2}^{(2)}&=\bra{\bf H}_{1,2}{\bf H}_{2,3}\ket,\\
{\rm K}_{1,2}^{(3)}&=\frac{1}{2!}\Big[\bra{\bf H}_{1,2}^2{\bf H}_{2,3}\ket+\bra{\bf H}_{1,2}{\bf H}_{2,3}^2\ket\Big],\\
{\rm K}_{1,2}^{(4)}&=\frac{1}{3!}\Big[\bra{\bf H}_{1,2}^3{\bf H}_{2,3}\ket+\frac{1}{2!}\big(2\bra{\bf H}_{1,2}^2{\bf H}_{2,3}^2\ket+\bra{\bf H}_{1,2}{\bf H}_{2,3}{\bf H}_{1,2}{\bf H}_{2,3}\ket\big)+\bra{\bf H}_{1,2}{\bf H}_{2,3}^3\ket\Big].
\end{align}
\end{subequations}
\item $m=3$
\begin{subequations}
\begin{align}
{\rm K}_{1,3}^{(3)}&=\bra{\bf H}_{1,2}{\bf H}_{2,3}{\bf H}_{3,4}\ket,\\
{\rm K}_{1,2}^{(4)}&=\frac{1}{2}\Big[\bra{\bf H}_{1,2}^2{\bf H}_{2,3}{\bf H}_{3,4}\ket+\frac{2}{3}\bra{\bf H}_{1,2}{\bf H}_{2,3}^2{\bf H}_{3,4}\ket+\bra{\bf H}_{1,2}{\bf H}_{2,3}{\bf H}_{3,4}^2\ket+\frac{1}{3}\bra{\bf H}_{1,2}{\bf H}_{2,3}{\bf H}_{3,4}{\bf H}_{2,3}\ket\Big].
\end{align}
\end{subequations}
\item $m=4$
\begin{subequations}
\begin{align}
{\rm K}_{1,4}^{(4)}&=\bra{\bf H}_{1,2}{\bf H}_{2,3}{\bf H}_{3,4}{\bf H}_{4,5}\ket.
\end{align}
\end{subequations}

\end{itemize}

\end{appendix}


\begin{thebibliography}{XX}
\bibitem{Jordan} The Jordan-Wigner transformation is given by:\\
 $\s_{m}^{-}={\rm e}^{i\pi\sum_{j=1}^{m-1}{\pmb a}_j^{\dag}{\pmb a}_{j}}{\pmb a}_m$, $\s_{m}^{+}=(\s_{m}^{-})^{\dag}$
with $\s_m^{\pm}=\frac{1}{2}(\s_m^{x}\pm i\s_m^{y})$ and $\s_{m}^{z}=2{\pmb n}_m-1$.
\end{thebibliography}
\end{document}